# Relation between the Usual Order and the Enumeration Orders of Elements of r.e. Sets

**Ali Akbar Safilian, Farzad Didehvar**

Department of Mathematics and Computer Science, Tehran Polytechnic University, Tehran, Iran

Emails: ali_safilian@aut.ac.ir, Didehvar@aut.ac.ir

## Abstract

In this paper, we have compared r.e. sets based on their enumeration orders with Turing machines. Accordingly, we have defined novel concept “uniformity” for Turing machines and r.e. sets and have studied some relationships between “uniformity” and both one-reducibility and Turing reducibility.

Furthermore, we have defined “type-2 uniformity” concept and studied r.e. sets and Turing machines based on this concept.

In the end, we have introduced a new structure called “Turing Output Binary Search Tree” that helps us lighten some ideas.

## 1 Introduction

Computability theory introduces partial functions $f{:}\subseteq N \to N$ by means of Turing machines. Let $A$ be an r.e. set. Infinite numbers of Turing machines produce the elements of $A$ and their enumeration orders may be different. Now assume that $A$ is a decidable set. We can design some Turing machines to produce the elements of $A$ in a desirable enumeration order. For instance, some Turing machines produce the elements of $A$ in ascendant usual order. Unlike decidable sets, this fact is not true for non-decidable r.e. sets.

In section 2, we introduce some new concepts to compare Turing machines and other concepts to compare r.e. sets based on their enumeration orders. In this section, a relation named “*uniformity*” is defined to compare r.e. sets and Turing machines. We prove that this relation is equivalence. In section 3, we found some relationships between “uniformity” and both one-reducibility and Turing reducibility. Some results in sections 2 and 3 motivated us to extend *uniformity* to *type-2 uniformity* in section 4. In this section, we obtained some results based on type-2 uniformity. Finally, in section 6, we introduce a novel structure named by *Turing Output Binary Search Tree* and show that it is a proper tool to show some ideas about enumeration orders of elements of r.e. sets.

We used the standard notions from recursion theory [6]. The set $\{1,2,\dots\}$ of natural numbers is denoted by $N$. The $n$th recursively enumerable set is denoted by $W_n$.

## 2 Uniformity on listings and sets

In this section, we define some concepts to compare different Turing machines and their generated sets upon their enumeration orders.

Let $A \subseteq N$ be an infinite non-empty recursively enumerable set. There are some total computable functions $h: N \to A$ such that $A = \{h(1), h(2), \dots\}$ (If the set $A$ is finite then $A = \{h(1), h(2), \dots, h(n)\}$ for $n = cardinality(A)$ and $h$ should be a partial function). In this paper, we call function $h$ a *listing* of $A$.

**Definition 2.1** A *listing* of an r.e. set $A \subseteq N$ is a bijective and surjective computable function $f: \subseteq N \to A$.

We can assign a listing to each Turing machine. Indeed a listing shows the *enumeration order* of output elements of the related Turing machine.

**Definition 2.2 (Uniformity on listings and sets)**

1. Two listings $h, g$ are uniform, $h \sim g$, if $h(i) < h(j) \Leftrightarrow g(i) < g(j)$ for all $i, j \in N$.
2. Two r.e. subsets $A, B$ of $N$ with equal cardinality are *uniform*, $A \sim B$, if there exist listings $h$ of $A$ and $g$ of $B$ such that $h \sim g$.

**Theorem 2.3** For r.e. sets $A, B$ with equal cardinality the following statements are equivalent:

1. $A \sim B$,
2. For every listing $h$ of $A$ there is a listing $g$ of $B$ such that $h \sim g$.

**Proof:**

**1⇒2:** Let $A \sim B$. Then there exist listings $h'$ of $A$ and $g'$ of $B$ such that $h' \sim g'$. Consider an arbitrary listing $h$ of $A$. We want to define a listing $g$ of $B$ such that $h \sim g$. Consider $g = g' o h'^{-1} o h$. It is evident that $g$ is a listing of $B$. We claim that $h \sim g$.

Consider $i, j \in N$ and $h(i) < h(j)$. Assume that $h(i) = a_1 \in A$ and $h(j) = a_2 \in A$. Since listings are surjective functions, there exist $m, k \in N$ such that $h'(k) = a_1$ and $h'(m) = a_2$.

$h(i) < h(j) \Rightarrow h'(k) < h'(m) \Rightarrow g'(k) < g'(m) \Rightarrow g'oh'^{-1}(a_1) < g'oh'^{-1}(a_2) \Rightarrow g'oh'^{-1}h(i) < g'oh'^{-1}h(j) \Rightarrow g(i) < g(j) \Rightarrow h \sim g$

2⇒1: it is evident…. □

**Lemma 2.4** The uniformity relation,$\sim$, on listings and sets is equivalence.

**Proof:** It is evident that uniformity on listings is an equivalence relation. This deduced directly from definition. Since the reflexivity and symmetric properties hold for "$\sim$" on listings, we can deduce easily that these properties hold for "$\sim$" on sets. Now it is sufficient to prove that the transitivity property holds for this relation on sets. Consider three sets $A, B, C$ such that $A \sim B$ and $B \sim C$. Then there exist two listings $h$ of $A$ and $g$ of $B$ such that $h \sim g$. According to lemma 2.3 there is a listing $f$ of $C$ such that $g \sim f$. Since transitivity holds for "$\sim$" on listings, we can deduce that $h \sim f$, so $A \sim C$. □

Let $A$ be an r.e. subset of $N$. We say $[A]_\sim = \{B \subseteq N \mid A \sim B\}$ is the "*uniformity equivalence class*" of $A$.

We illustrate the above concepts by the following example.

**Example 2.5** Consider two listings $h$ of the set $A = \{2i : i \in N\}$ and $g$ of the set $B = N - \{1\}$ such that for all $i \in N$, $h(i) = 2i$ and $g(i) = i + 1$. It is clear that for all $i, j \in N$, $h(i) < h(j) \Leftrightarrow g(i) < g(j)$, so $h \sim g$. Therefore, $A \sim B$.□

**Lemma 2.6** Two finite sets with equal cardinality are uniform.

**Proof:** Consider two sets $A, B$ are finite subsets of $N$ with equal cardinality**.** Since every finite subset of $N$ is recursive, there exist listings $h$ of $A$ and $g$ of $B$ such that for all $i, j \in dom(h)$, $i < j \Leftrightarrow h(i) < h(j)$ and $for\ all\ \ i, j \in dom(g), i < j \Leftrightarrow g(i) < g(j)$. Cardinality of these two sets is equal and thus these two listings are uniform. Hence, two sets $A, B$ are uniform. □

In the following, we introduce two sets that are not uniform. Henceforth, we consider $K = \{ n \in N \mid n \in W_n \}$ as usual.

**Lemma 2.7** Two sets $N$ and $K$ are not uniform.

**Proof**: For the sake of a contradiction, assume that these two sets are uniform. The identity function $id: N \to N$ is a listing of $N$. Then, according to lemma 2.3, there exists a listing $g$ of $K$ such that $h \sim g$. Therefore, for all $i, j \in N,\ i < j \Leftrightarrow g(i) < g(j)$. But this cannot be true. This is a contradiction. □

## 3 One-reducibility, Turing-reducibility and uniformity

In this section, we want to explore some relationships between both one-reducibility & Turing-reducibility and uniformity on sets. First, we investigate one-reducibility equivalence classes.

**Lemma 3.1** Consider a non-decidable r.e. set $A$ such that there is not any Turing machine to obtain minimum element of it. Two sets $A$ and $B = A \cup \{1\}$ are not uniform.

**Proof:** For the sake of a contradiction, assume that these two sets are uniform. Consider a listing $h$ of $B$ such that $h(1) = 1$. Then there exists a listing $g$ of $A$ such that $h \sim g$. Since $h(1)$ is the minimum element of $B$ then $g(1)$ should be the minimum element of $A$. This shows that we could compute the minimum element of $A$. It leads us to Contradiction. □

**Lemma 3.2** Two sets $A = \{2i : i \in N\}$ and $B = N - \{1\}$ do not belong to the same one-reducibility equivalence class.

**Proof**: For the sake of a contradiction, assume that these two sets belong to the same one-reducibility equivalence class. In the sense of the definition of one-reducibility, two sets $A^c$ and $B^c$ are of equal cardinality. $A^c$ is an infinite set and in contrast, the cardinality of $B^c$ is 1. This is a contradiction. □

**Proposition 3.3** If two r.e. sets belong to same "one-reducibility equivalence class", then they do not belong necessarily to same "uniformity equivalence class".

**Proof:** Consider a non-decidable r.e. set $A$ such that there is not any Turing machine for obtain minimum element of it. In the lemma 3.1, we proved that $A$ and $A \cup \{1\}$ are not uniform, whereas these sets belong to on-reducibility equivalence class $[A]_{1-reducibility}$. □

**Proposition 3.4** If two r.e. sets belong to the same "uniformity equivalence class", then they do not belong necessarily to the same "one-reducibility equivalence class". □

**Proof:** In the example 2.5, we showed that two sets $A = \{2i : i \in N\}$ and $B = N - \{1\}$ are uniform and in the lemma 3.2, we showed that they do not belong to the same one-reducibility equivalence class.

Now in the sequel to this section we investigate some relationships between Turing-reducibility and uniformity.

For any r.e. sets $A$ and $B$ if $\mathrm{A} \equiv_1 \mathrm{B}$ then $\mathrm{A} \equiv_{\mathrm{T}} \mathrm{B}$. Therefore, the following corollary is deduced from the proposition 3.3.

**Corollary 3.5** If two r.e. sets belong to the same Turing-reducibility equivalence class, then they do not belong necessarily to the same uniformity equivalence class. □

In the following, we prove that if two sets belong to the same uniformity equivalence class then they belong to the same Turing-reducibility equivalence class.

**Theorem 3.6** If two r.e. sets $A$ and $B$ are uniform then they belong to the same Turing-reducibility equivalence class.

**Proof**: If two sets $A$ and $B$ are uniform then for any listing $h$ of $A$ there exists a listing $g$ of $B$ such that $h \sim g$. Therefore, for all $i, j \in N, h(i) < h(j) \Leftrightarrow g(i) < g(j)$.This shows that $B$ is recursive relative to $A$ and symmetrically $B$ is recursive relative to $A$. Therefore, $\mathrm{A} \equiv_{\mathrm{T}} \mathrm{B}$ . □

Theorem 3.6 shows that for any r.e. set $A$, $[\mathrm{A}]_{\sim} \subseteq [\mathrm{A}]_{\mathrm{T}}$. Now we want to prove that for any r.e. set $A \subseteq N$ there are infinite numbers of uniformity equivalence classes such that they are subsets of the equivalence class $[\mathrm{A}]_T$.

**Theorem 3.7** Let $A$ be an r.e. set. There are infinite numbers of r.e. sets $B$ such that $[\mathrm{B}]_{\sim} \subseteq [\mathrm{A}]_T$.

**Proof:** There exist two cases:

CASE 1: $A$ is a decidable set.

According to the lemma 2.6, all recursive sets of equal cardinality belong to the same uniformity equivalence class. We know that every recursive set is a member of the set $[\emptyset]_T$. Since all uniform sets are of equal cardinality, there exist infinite numbers of uniformity equivalence classes such that they are subsets of the equivalence class$[\emptyset]_T$.

CASE 2: $A$ is a non-decidable r.e. set. We know that there is a set $C \in [A]_T$ such that there is not any Turing machine to obtain minimum element of it.

In the similar way as used in the lemma 3.1, we can deduce that any two distinct members of the set series $(\mathrm{C}^{\mathrm{i}} = \mathrm{C} \cup \{1,2,\ldots,\mathrm{i}\})_{\mathrm{i} \in \mathrm{N}}$ are not uniform. Based on the theorem 3.6, we know that for any $i \in N$, $[\mathrm{C}^{\mathrm{i}}]_{\sim} \subseteq [\mathrm{A}]_T$. Therefore, there are infinite numbers of uniformity equivalence classes such that they are subsets of the equivalence class $[\mathrm{A}]_T$. □

## 4 Type-2 uniformity

Adding some finite numbers of elements to some non-decidable r.e. sets, we obtained the non-uniform sets. You can see this in some previous lemmas and theorems such as lemma 3.7. This fact motivates us to define type-2 uniformity relation.

**Definition 4.1** We say **t**wo sets $A, A'$ are *almost equal*, $A \approx A'$ if $A \nabla A'((A - A') \cup (A' - A))$ is a finite set.

It is clear that $\approx$ is an equivalence relation.

**Definition 4.2 (Type-2 uniform sets)**

We say two sets $A$ and $B$ are type-2 uniform if there exist $A', B'$ such that $A \approx A'$ , $B \approx B'$ and $A' \sim B'$.

Since two finite sets of equal cardinality are type-2 uniform, henceforth we consider only infinite r.e. sets.

**Lemma 4.3** Two sets $A, B$ are type-2 uniform, $A \sim_2 B$, if, and only if, there exist finite sets $C, D$ such that $A - C \sim B - D$.

**Proof:** Consider $A \sim_2 B$. In the sense of the definition 4.2, there exist two finite sets $A'' = \{a''_1, \dots, a''_n\}$ and $B'' = \{b''_1, \dots, b''_m\}$ such that one of the following cases is true:

1. $(A - A'') \sim (B - B'')$,
2. $(A - A'') \sim (B \cup B'')$,
3. $(A \cup A'') \sim (B - B'')$,
4. $(A \cup A'') \sim (B \cup B'')$.

Since the cases 2 and 3 are similar, we consider only two cases 2, and 4.

CASE 2: Let $h$ be a listing of $B$. We define a listing $f$ of $(B \cup B'')$ as follows:

$$f(i) = \begin{cases} b''_i & \text{if } 1 \le i \le m \\ h(i - m) & \text{otherwise} \end{cases}$$

Since two sets $(A - A'')$ and $(B \cup B'')$ are uniform, there exists a listing $t$ of $(A - A'')$ such that $f \sim t$.

We define a listing $l$ of $A - A'' - \{t(1), \dots, t(m)\}$ as follows: for all $i \in N$ $\quad l(i) = t(i + m)$. It is evident that listings $l$ and $h$ are uniform, so two sets $A - A'' - \{t(1), \dots, t(m)\}$ and $B$ are uniform sets. We could reduce the case 2 to the case 1.

CASE 4: Let $h$ be a listing of $B$. We define a listing $f$ of $(B \cup B'')$ as follows:

$$f(i) = \begin{cases} b''_i & \text{if } 1 \leq i \leq m \\ h(i-m) & \text{otherwise} \end{cases}$$

Since two sets $(A \cup A'')$ and $(B \cup B'')$ are uniform, there exists a listing $t$ of $(A \cup A'')$ such that $f \sim t$.

Let $p \in N$ be the minimum number such that for all $i \in N$, $t(i) \in \{a''_1, \dots, a''_n\} \Rightarrow i \leq p$. Consider two sets $C = A \cup A'' - \{t(1), \dots, t(\max(p,m))\}$ and $D = B \cup B'' - \{f(1), \dots, f(\max(p,m))\}$. It is clear that $A'' \subseteq \{t(1), \dots, t(\max(p,m))\}$ and $B'' \subseteq \{f(1), \dots, f(\max(p,m))\}$. Therefore, two sets $A' = \{t(1), \dots, t(\max(p,m))\} - A''$ and $B' = \{f(1), \dots, f(\max(p,m))\} - B''$ are finite. Since $(A \cup A'') \sim (B \cup B'')$ and $f \sim t$ , the two sets $A - A'$and $B - B'$ are uniform too. We could reduce the case 2 to the case 1. □

**Definition 4.4** Two listings $h, g$ are called type-2 uniform, $h \sim_2 g$, if there exist $m, n \in N$ such that two uniform listings $h', g'$ are defined as follows: for all $i \in N$ $h'(i) = h(i+m), g'(i) = g(i+n)$. We say that $m$ and $n$ *satisfy* $h$ and $g$ respectively to be type-2 uniform.

**Lemma 4.5** For infinite r.e. sets $A, B$ the following statements are equivalent:

1. $A \sim_2 B$,
2. There exist listings $h$ of $A$ and $g$ of $B$ such that $h \sim_2 g$,
3. For any listing $h$ of $A$ there exists a listing $g$ of $B$ such that $h \sim_2 g$.

**Proof:**

**1⟹2:** Assume that infinite r.e. sets $A, B$ are type-2 uniform. According to the lemma 4.3, there exist finite sets $A' = \{a'_1, \dots, a'_n\}, B' = \{b'_1, \dots, b'_m\}$ such that $(A - A') \sim (B - B')$. This means that there exist listings $h$ of $(A - A')$ and $g$ of $(B - B')$ such that $h \sim g$. We define a listing $f$ of $A$ and $t$ of $B$ as follows:

$$f(x) = \begin{cases} a'_x, & 1 \leq x \leq n \\ h(x-n) & x > n \end{cases}$$

$$t(x) = \begin{cases} b'_x, & 1 \leq x \leq m \\ g(x-n) & x > m \end{cases}$$

Therefore, for all $i \in N$, $h(i) = f(i+n)$ and $g(i) = t(i+m)$. Since, two listings $g, h$ are type-2 uniform, $f \sim_2 t$.

**2⟹1:** Assume that two listings $h$ of $A$ and $g$ of $B$ are type-2 uniform. Then there are $m, n \in N$ such that satisfy $h$ and $g$ respectively to be type-2 uniform. Therefore, two sets $A-\{h(1),\dots,h(m)\}$ and $B-\{g(1),\dots,g(n)\}$ are uniform. This shows that two sets $A$ and $B$ are type-2 uniform, $A\sim_2 B$.

**2⟹3:** Assume that two listings $h$ of $A$ and $g$ of $B$ are type-2 uniform and two numbers $m,n$ satisfy $h$ and $g$ respectively to be type-2 uniform. Therefore, two sets $A-\{h(1),\dots,h(m)\}$ and $B-\{g(1),\dots,g(n)\}$ are uniform. Consider that $f$ is a listing of $A$. Let $p\in N$ be the minimum number such that for all $i\in N$ $f(i)\in\{h(1),..,h(m)\}\Rightarrow i\leq p$. Since two sets $A-\{h(1),\dots,h(m)\}$ and $B-\{g(1),\dots,g(n)\}$ are uniform and $\{h(1),\dots,h(m)\}\subseteq\{f(1),\dots,f(p)\}$, two sets $A-\{f(1),\dots,f(p)\}$ and $B-\{g(1),\dots,g(n+p-m)\}$ are uniform too. We define a listing $f'$ as follows: for all $i\in n$ $f'(i)=f(i+p)$. $f'$ is a listing of $A-\{f(1),\dots,f(p)\}$ so there is a listing $t'$ of $B-\{g(1),\dots,g(n+p-m)\}$ such that $f'\sim g'$. Now we define a listing $t$ of $B$ as follows:

$$t(x)=\begin{cases} g(x), & x\leq n+p-m \\ t'(i-n-p+m), & x>n+p-m \end{cases}$$

Numbers $p$ and $(n+p-m)$ satisfy listings $f$ and $t$ respectively to be type-2 uniform. Therefore, for any listing $f$ of $A$ there exists a listing $t$ of $B$ such that $f\sim_2 t$.

**3⟹2:** It is evident. □

**Lemma 4.6** The type-2 uniformity $\sim_2$ relation on both listings and sets is an equivalence relation.

**Proof:** Assume that two listings $h$ and $g$ are type-2 uniform. There are $m,n\in N$ such that satisfy $h,g$ respectively to be type-2 uniform. Therefore, two sets $A-\{h(1),\dots,h(m)\}$ and $B-\{g(1),\dots,g(n)\}$ are uniform. According to the lemma 2.4, uniformity on sets is an equivalence relation and thus type-2 uniformity on listings is an equivalence relation too.

Now assume that two sets $A,B$ are type-2 uniform. According to the lemma 4.5, there exist listings $h$ of $A$ and $g$ of $B$ such that $h\sim_2 g$. Since type-2 uniformity on listings is an equivalence relation, $\sim_2$ on sets is an equivalence relation too. □

We denote type-2 uniformity equivalence class of an r.e. set $A$ by $[A]_{\sim_2}=\{B\subseteq N\mid A\sim_2 B\}$.

**Notation:** In the subsequent lemmas and theorems, we utilize the following notations for every two listings $h$ and $g$:

1) For all $m,n\in N$ the notation $E_{m,n}^{h,g}$ denotes the set $\{(i,j)\in N^2\mid h(i+m)<h(j+m)\ and\ g(i+n)>g(j+n)\}$,

2) For all $m, n \in N$ the notation $M_{m,n}^{h,g}$ denotes the set
$\{i \in N |\ there\ exists\ a\ j \in N\ such\ that\ (i,j) \in E_{m,n}^{h,g}\}$.
3) For all $m, n \in N$ the notation $L_{m,n}^{h,g}$ denotes the set
$\{j \in N |\ there\ exists\ a\ i \in N\ such\ that\ (i,j) \in E_{m,n}^{h,g}\}$.

**Lemma 4.7** If listings $h, g$ are not type-2 uniform, then for all $m, n \in N$, $M_{m,n}^{h,g}$ and $L_{m,n}^{h,g}$ are infinite.

**Proof:** For the sake of a contradiction, assume that $L_{m,n}^{h,g}$ and $M_{m,n}^{h,g}$ are finite. Now assume that $a = max\{i | i \in M_{m,n}^{h,g}\}$ and $b = \max\{j | j \in L_{m,n}^{h,g}\}$. It is clear that two numbers $(a + m)$ and $(b + n)$ satisfy listings $h, g$ to be type-2 uniform. This leads us to a contradiction. □

**Lemma 4.8** Let r.e. sets $A, B$ are not type-2 uniform. Consider an r.e. set $C$ such that $A \nsubseteq C$. $A \cup C$ and $B$ are not type-2 uniform.

**Proof:** It is evident that the set of all listings of an r.e. set is enumerable. Assume that the listings of $B$ are $\{g_1, g_2, \ldots\}$. Let $l$ and $h$ be listings of $C - A$ and $A$ respectively.

We define sequences $\{S_n\}_{n \in N}$ , $\{E_n\}_{n \in N}$ and $\{I_n\}_{n \in N}$ as follows:

$$S_n = \begin{cases} 1 & if\ n = 1 \\ E_{n-1} + n & otherwise \end{cases}$$

$$E_n = S_n * 10$$

$$I_n = E_n - S_n$$

Now we want to define a listing $f$ of $A \cup B$ recursively based on the $h, l$ as follows:

$$f(x) = \begin{cases} h(x) & if\ S_n \le x \le E_n \\ l(n) & if\ x = E_n + 1 \\ h(E_{n-1} + x - E_n - 1) & if\ E_n + 2 \le x < S_{n+1} \end{cases}$$

The following figure shows thoroughly the construction of $f$.

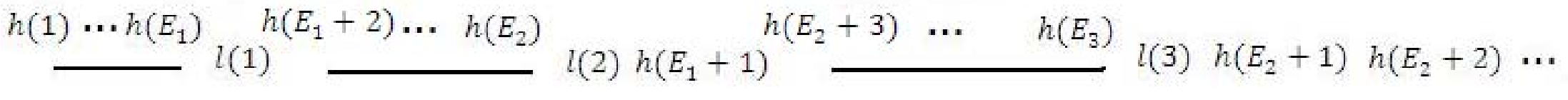

Figure 4.1

Since $A$ and $B$ are not type-2 uniform, for all $i \in N$ listings $h$ and $g_i$ are not type-2 uniform and thus according to the lemma 4.7, $L_{m,n}^{h,g_i}$ and $M_{m,n}^{h,g_i}$ are infinite.

We claim that for all $i \in N$, $f$ and $g_i$ are not type-2 uniform. For the sake of a contradiction, assume that there exist $k, m, n \in N$ such that for all $i, j \in N$ $f(i+m) < f(j+m) \Leftrightarrow g_k(i+n) < g_k(j+n)$. [1]

Since the sequence $\{I_n\}_{n\in N}$ is ascendant, and $L^{h,g_k}_{m,n}, M^{h,g_k}_{m,n}$ are infinite, there are $p, i, j \in N$ such that $S_p \leq i+m, j+m \leq E_p$ and $(i,j) \in E^{h,g_k}_{m,n}$. Since $h(i+m) = f(i+m)$ and $h(j+m) = f(j+m)$, $f(i+m) < f(j+m) and\ g_k(i+n) < g_k(j+n)$. [2]

[1] and [2] lead to a contradiction. □

In section 3, we argued that if two sets belong to the same "one-reducibility equivalence class", then they do not belong necessarily to the same "uniformity equivalence class". The lemma 3.1 supported it. In the following, we prove that this fact is true for type-2 uniformity too.

**Lemma 4.9** If two sets belong to the same one-reducibility equivalence class, then they are not necessarily type-2 uniform.

**Proof:** Let $A$ be a non-decidable r.e. set and $B$ an infinite recursive set such that $A \nsubseteq B$. We can clearly say that two sets $A$ and $A \cup B$ belong to the same one-reducibility equivalence class but according to the lemma 4.7, they are not type-2 uniform. □

Furthermore, in the section 3 we argued that two uniform sets do not belong necessarily to same "one-reducibility equivalence class". Since two uniform sets are type-2 uniform, we can extend simply proposition 3.4 to type-2 uniformity.

**Proposition 4.10** Two type-2 uniform sets do not belong necessarily to same one-reducibility equivalence class.

Like of the corollary 3.5, we can deduce that two r.e. sets which belong to same Turing-reducibility equivalence class, are not necessarily type-2 uniform.

Now, we want to survey Theorem 3.6 by type-2 uniformity. This theorem says that if two r.e. sets $A$ and $B$ are uniform then they belong to same Turing-reducibility equivalence class. We prove in the following that this is true for type-2 uniformity too.

**Theorem 4.11** If two r.e. sets $A$ and $B$ are type-2 uniform then they belong to same Turing-reducibility equivalence class.

**Proof:** $A$ and $B$ are type-2 uniform and thus according to the lemma 4.3, there are two finite sets $A' \subseteq A$ and $B' \subseteq B$ such that two sets $A'' = A - A'$ and $B'' = B - B'$ are uniform. According to the theorem 3.6, two sets $A'', B''$ belong to same Turing-reducibility equivalence class. Since $A'$ and $B'$ are recursive and $[A'']_T = [B'']_T$, two sets $A, B$ belong to same Turing-reducibility equivalence class. □

At last, we want to survey theorem 3.7. This theorem says that each Turing-reducibility equivalence class has infinite subsets of uniformity equivalence classes. This is not true for type-2 uniformity. Because in the following, we prove that $[\emptyset]_T$ contain only two type-2 uniformity equivalence class.

**Lemma 4.12** There are only two type-2 uniformity equivalence classes such that they are subsets of the equivalence class $[\emptyset]_T$.

**Proof:** It is clear that every two finite sets are type-2 uniform even with different cardinalities but a finite set and an infinite recursive set are not type-2 uniform. This shows that there are two numbers of uniformity equivalence classes such that they are subsets of the equivalence class$[\emptyset]_T$ . One of them is the set of all infinite recursive sets and another is the set of finite sets. □

Now, consider a non-decidable r.e. set $A$. According to the lemma 4.11, $[\mathrm{A}]_{\sim_2} \subseteq [\mathrm{A}]_T$. How many type-2 uniformity equivalence classes are subsets of the $[\mathrm{A}]_T$? Consider an infinite recursive set $B \not\supseteq A$. Based on the lemma 4.8, two sets $A$ and $C = A \cup B$ are not type-2 uniform sets. They belong to same Turing reducibility equivalence class $[\mathrm{A}]_T$, i.e. $[\mathrm{C}]_{\sim_2} \subseteq [\mathrm{A}]_T$. Therefore, for every recursively enumerable non-decidable set $A$, there are at least two type-2 uniformity equivalence classes subsets of the $[\mathrm{A}]_T$.

**Theorem 4.13** For every non-decidable r.e. set $A$, there are infinite numbers of type-2 uniformity equivalence classes such that they are subsets of equivalence class$[\mathrm{A}]_T$.

**Proof:** Since recursively enumerable Turing degrees are dense, there is a chain of non-decidable r.e. sets $\{A_i\}_{i \in N}$ such that for all $i \in N$ , $A_{i+1} <_T A_i$ and $A_0 = A$ and for distinct members of this sequence $C$ and $D$, $C \nsubseteq D$ . For all $i, j \in N$, $[A_i]_T \neq [A_j]_T$ and thus based on the theorem 4.11, $[A_i]_{\sim_2} \neq [A_j]_{\sim_2}$. Consider the increasing chain $\{B_i = \bigcup_{0 \leq k \leq i} A_k\}_{i \in N}$. For all $i \in N$, $B_i \in [A]_T$. According to the lemma 4.8, for all $i, j \in N \; and \; i \neq j$, $B_i$ and $B_j$ are not type-2 uniform. Then there are infinite numbers of type-2 uniformity equivalence classes such that they are subsets of the equivalence class $[\mathrm{A}]_T$. □

## 5 Turing Output Binary Search Tree

In computer science, a *binary search tree (BST)* is a binary tree data structure. In this structure, each node has a value and a total order is defined on these values. The left subtree of a node contains only values less than the node's value; the right subtree of a node contains only values greater than or equal to the node's value. The major advantage of binary search trees over other data structures is that the related *sorting algorithms* and *search algorithms* such as *in-order traversal* can be very efficient. Binary search trees are a fundamental data structure used to construct more abstract data structures such as sets, multi-sets, and associative arrays. [5]

We designed a new structure named "*Turing Output Binary Search Tree*" (abbreviated by "*TOBST*") that is a binary search tree constructed for a given Turing machine on the basis of enumeration orders of its produced elements.

Consider a Turing machine $M$ with related listing $h: \subseteq N \to A$. This means this machine produces the elements of $A$. For any $i \in N$ the notation $\mathrm{TOBST}^{i}_{(h,A)}$ denotes the construction result in step $i$. Note that the nodes of each TOBST are couple. For example, the node $(5, h\ (5))$ shows that the fifth output element of $M$ is $h(5)$. In the following, we describe construction procedure of TOBST for an r.e. set $A$ with the related listing $h$.

> $\mathrm{TOBST}^{0}_{(h,A)}$ is an empty tree.
> $\mathrm{TOBST}^{1}_{(h,A)}$ is a tree with only one node $(1, h(1))$.
> $\mathrm{TOBST}^{n}_{(h,A)}$ is constructed based on $\mathrm{TOBST}^{n-1}_{(h,A)}$ as follows:
> Insertion $(n, h(n))$ to $\mathrm{TOBST}^{n-1}_{(h,A)}$ to obtain the $\mathrm{TOBST}^{n}_{(h,A)}$ is our aim and begins as search would begin; we examine the value of root and recursively insert the new node to the left sub tree if the new value $h(n)$ is less than the value of root, or the right sub tree the new value is greater than the value of root. (Note for all $i \in N$ the value of the node $(i, h(i))$ is $h(i)$).

The notation $\mathrm{A}^{\mathrm{i}}_{\mathrm{h}}$ denotes the set $\mathrm{A}^{\mathrm{i}}_{\mathrm{h}} = \{h(i) : \mathrm{i} \in \{1,2, \dots, i\}\}$.

We illustrate the concept of TOBST by the following example.

**Example 5.1** Consider $A = \{a_1, a_2, a_3, a_4, a_5\}$ such that for all $i, j \in \{1,2,3,4,5\}$ $i < j \leftrightarrow a_i < a_j$. In the following, we introduce three different Turing machines, which produce the elements of the set $A$ and show that TOBST of each of these machines are different.

Consider $h$: $\{1, \dots, 5\} \to A$ is a listing of the set $A$ in which $h(i) = a_i$, for all $i \in dom(h)$. Therefore, the usual order of outputs is ascendant. In the following, we show the construction procedure of its TOBST step by step;

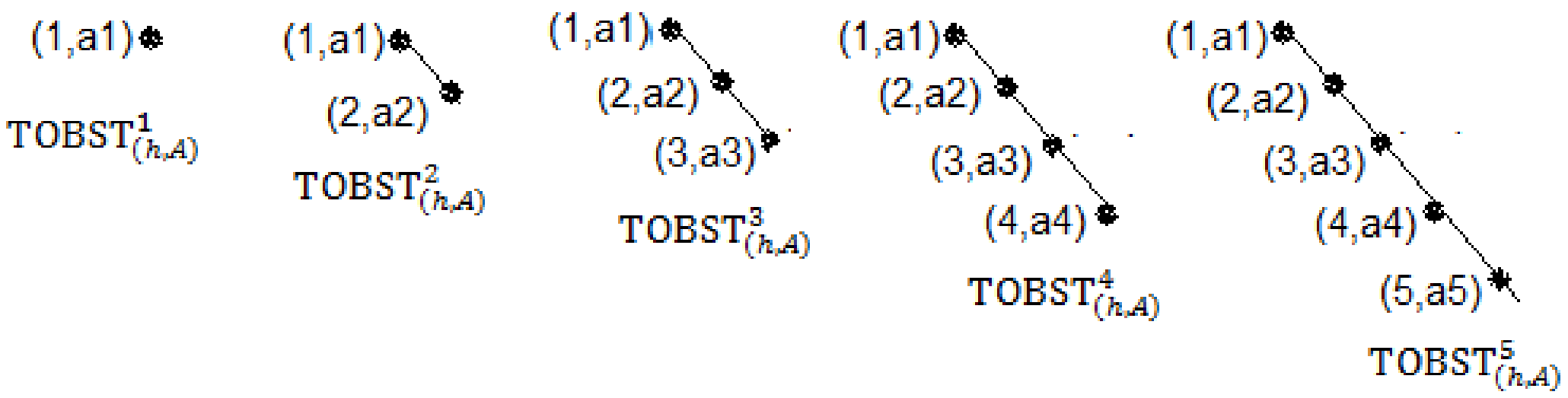

Figure 5.1

Consider $h: \{1, \dots, 5\} \rightarrow A$ is a listing of the set $A$ in which $h(i) = a_{6-i}$, for all $i \in dom(h)$. The construction of its TOBST showed as follows;

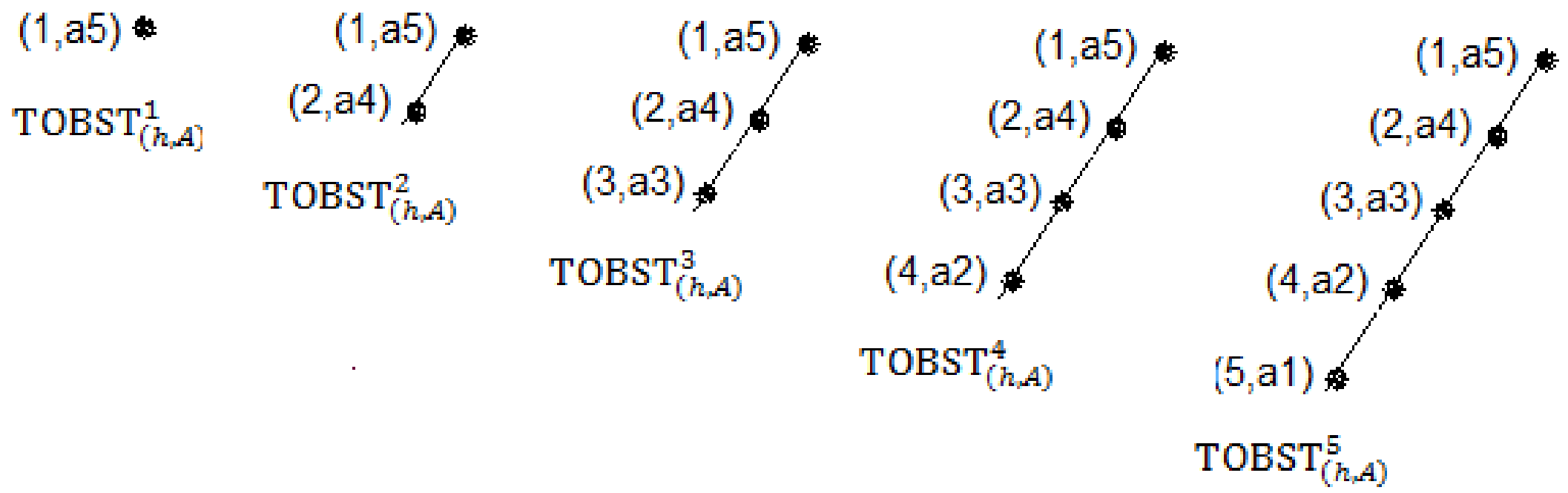

Figure 5.2

Consider $h: \{1, \dots, 5\} \rightarrow A$ is the production index function of the set $A$ in which $h(1) = a_3$, $h(2) = a_5$, $h(3) = a_4$, $h(4) = a_1$, $h(5) = a_2$. The construction of its TOBST showed as follows;

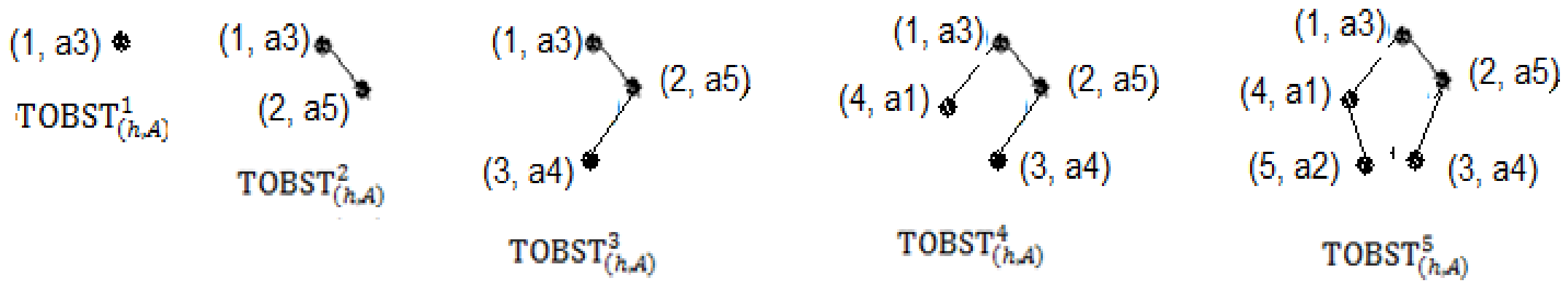

Figure 5.3

A set $A \subseteq N$ is recursive if, and only if, it has a monotonic listing. Based on this simple fact, it supports that $A$ is decidable if, and only if, it has a listing $h$ such that for all $i \in dom(h)$, TOBST$^i_{(h,A)}$ is a right crisscross tree.

Consider a listing $h$ of an r.e. set $A \subseteq N$ such that for all $i \in dom(h)$, TOBST$^i_{(h,A)}$ is a left crisscross tree. Since in this case $h(1)$ must be the maximum element of $A$ , $A$ is a finite set.

We claim that TOBST structure extends our ability to define useful concepts.

**Definition 5.2 (Isomorphic listings in step i)** Let $A$ and $B$ be r.e. sets. We called Listings $h$ of $A$ and $g$ of $B$ *isomorphic in step i* if TOBST$^i_{(h,A)}$ and TOBST$^i_{(g,B)}$ are isomorphic trees.

We illustrate the above definition by following examples. First example shows two isomorphic listings that their related Turing machines produce the elements of a similar set and second example shows two listings that their related Turing machines produce different sets.

**Example 5.3** Consider $B = \{1, 2, 5, 6, 7, 8, 14\}$ and a recursively enumerable set $A$ such that $B \subseteq A$ and two Turing machines $M_1$ and $M_2$ such that both produce the elements of $A$ with assigned listings $h_1$ and $h_2$ respectively. Assume that $h_1(1) = 7, h_1(2) = 2, h_1(3) = 5, h_1(4) = 6, h_1(5) = 14$; $and$ $h_2(1) = 6, h_2(2) = 8, h_2(3) = 1, h_2(4) = 2, h_2(5) = 5$. The figures 5.4 and 5.5 show the construction procedure of TOBST$^5_{(h1,A)}$ and TOBST$^5_{(h2,A)}$ respectively.

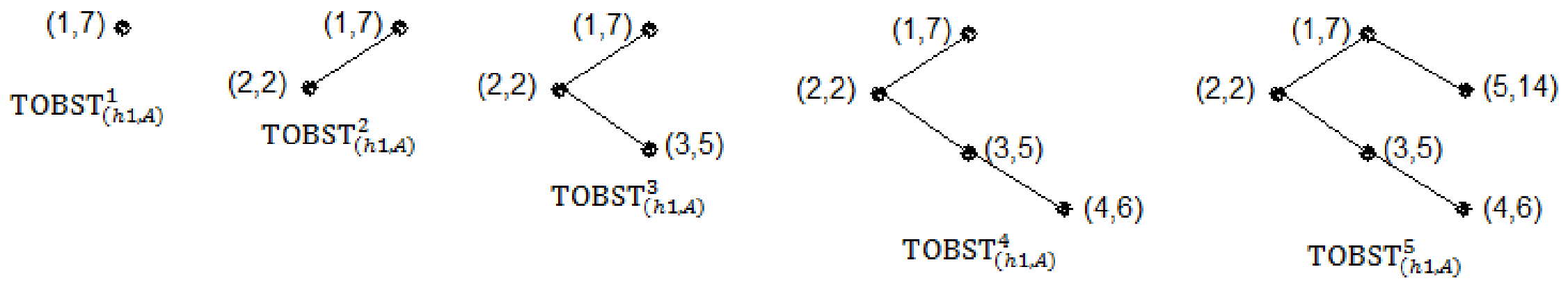

Figure 5.4

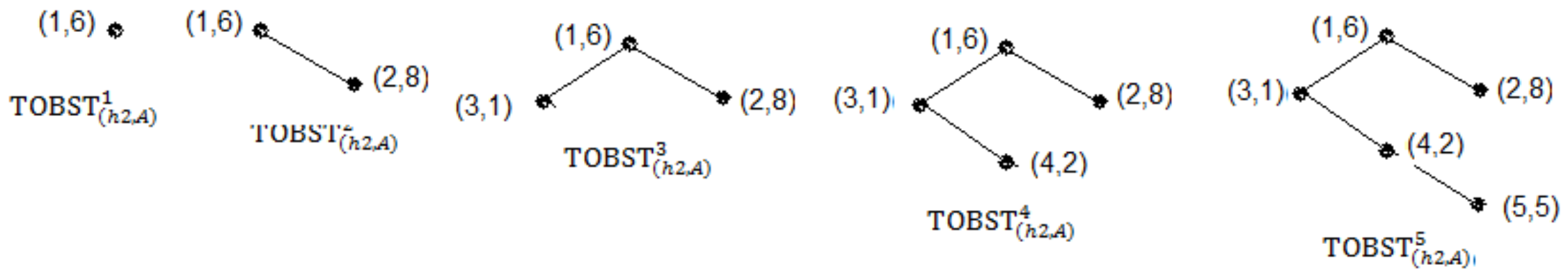

Figure 5.5

As you see, two trees TOBST$^5_{(h1,A)}$ and TOBST$^5_{(h2,A)}$ are isomorphic, so two listings $h_1$ and $h_2$ are isomorphic in step 5. In addition, you can see that for any $1 < i < 5$ TOBST$^i_{(h1,A)}$ and TOBST$^i_{(h2,A)}$ are not isomorphic.

**Example 5.4** Let two sets $A$ and $B$ be recursive such that $\{7,2,5,6,14\} \subseteq A$ and $\{6,8,1,2,5\} \subseteq$ B. Consider two listings $h$ of $A$ and $g$ of $B$ such that $h(1) = 7, h(2) = 2, h(3) = 5, h(4) = 6, h(5) = 14$ and $g(1) = 6, g(2) = 8, g(3) = 1, g(4) = 2, g(5) = 5$. It is clear that TOBST$^5_{(h,A)}$ and TOBST$^5_{(g,B)}$ are isomorphic trees. Therefore $h$ and $g$ are isomorphic in step 5. □

TOBST is a proper tool for visualizing relationship between usual order and enumeration orders in Turing machines and help us to show some ideas properly and define new concepts. By this concept, we can transfer all the concepts mentioned in the previous sections such as uniformity to trees. For instance in the following we transfer the concept of uniformity on listings. Other concepts can transfer to trees easily.

**Proposition 5.5 (uniform listings with TOBST)** two listings $h$ of $A$ and $g$ of $B$ are *uniform* if, and only if, for all $i \in N$, $\text{TOBST}^{i}_{(h,A)}$ and $\text{TOBST}^{i}_{(g,B)}$ are isomorphic trees.

We can replicate this transferring for other introduced concepts about uniformity. On the other hand, we can introduce some valuable concepts by TOBST such as the definition 5.2.

## Acknowledgment

The authors wish to thank Professors Klaus Weihrauch and Marat Arslanov for their valuable sympathy and precious hints to prepare this paper.